\documentclass[11pt,draftclsnofoot, journal, onecolumn]{IEEEtran}
\usepackage{amsmath,cite,amsfonts,amssymb,psfrag}
\usepackage{graphicx}
\usepackage{rotating}
\usepackage{color}

\newtheorem{theorem}{Theorem}

\newtheorem{remark}{Remark}
\newtheorem{definition}{Definition}

\newtheorem{lemma}{Lemma}
\newtheorem{corollary}{Corollary}

\begin{document}
\title{Secret Key Generation Through a Relay} 
\IEEEoverridecommandlockouts
\author{
\authorblockN{Kittipong Kittichokechai, Rafael F. Schaefer, and Giuseppe Caire\\
\authorblockA{Technische Universit\"{a}t Berlin}
}
}
\maketitle
\begin{abstract}
 We consider problems of two-user secret key generation through an intermediate relay. 
 Each user observes correlated source sequences and communicates to the relay over rate-limited noiseless links. The relay processes and broadcasts  information to the two users over another rate-limited link. The users then generate a common secret key based on their sources and  information from the relay. 
 In the untrusted relay setting, the goal is to establish  key agreement between the two users at the highest key rate without leaking information about the key to the relay. We characterize inner and outer bounds to the optimal tradeoff between communication and key rates. The inner bound is based on the scheme which involves a combination of binning, network coding, and key aggregation techniques. For the trusted relay setting with a public broadcast link, the optimal communication-key rate tradeoff is provided for a special case where the two sources are available losslessly at the relay. 
The results can be relevant for cloud-based secret key generation services.
\end{abstract}
\section{Introduction}\label{sec:introduction}
Cloud-based services have gained significant interests 
with a growing adoption for both personal and business uses. 
The main idea involves shifting  computational tasks traditionally done at user's devices to the cloud server/processor which is accessible through some communication channels. 
This approach has enabled several functionalities especially for small and less powerful devices. 
Despite its usefulness, serious concerns regarding information security and privacy arise due to the fact that information available at the cloud server could be undesirably exploited. 

In this work, we consider one particular cloud-based
service, namely secret key generation of two users. The two users have individual access to two correlated sources and 
communicate separately to the cloud server (the relay) who then processes received information and sends back a common broadcast message
to the received users to complete the key generation process. This type of cloud-based services is relevant in scenarios where the communication between users 
must occur through some network infrastructure. For example, in today's Internet, the communication between two users occurs typically through routers. Our model applies immediately to such a scenario, by forcing the relay to simply rebroadcast the two messages sent by the users. Including some computing capability at the relay (namely, ``cloud computing''), the tradeoff between communication and common key rate can be improved. However, when the relay is involved in the key generation process, it can also gain knowledge of the users' common key, which is then no longer secure. 
This motivates our problem formulation in terms of   secret key generation through an untrusted relay, as shown in Fig. 1. The security constraint is reflected by the requirement that the information leakage  rate at the relay should be kept arbitrarily small. 

 In this work, we characterize inner and outer bounds to the optimal tradeoff region of communication rates and key rate. The inner bound is based on an achievability scheme which involves a novel combination of binning, network coding, and key aggregation techniques. 

It is also interesting to consider the case where the relay is \emph{trustworthy} but its broadcast communication to the users is over a public channel, as shown in Fig. \ref{fig:model2}. In this case, we provide a complete characterization of the communication-key rate region for a special case where the two sources are available directly at the relay.

Our problem is closely related to works on secret key agreement over public communications, introduced in \cite{acCRIT93}, \cite{mSKAP93}. 
It is also related to the problem of secret key generation with a helper \cite{cnCRSK00} where the helper provides rate-limited side information to both users. 
In our problem, the broadcast information from the relay may be considered as  helper information. However, this information is a function of users' own information rather than that of another correlated source at the helper. In this sense, the key generation process involves some ``feedback" information. Other related works on multi-terminal secret key generation include, e.g., \cite{cnSCFM04,tCIAS13,lcvCRKG16, ssaRRSK11}.  
\begin{figure}[]
	\centerline{\includegraphics[width=9.5cm]{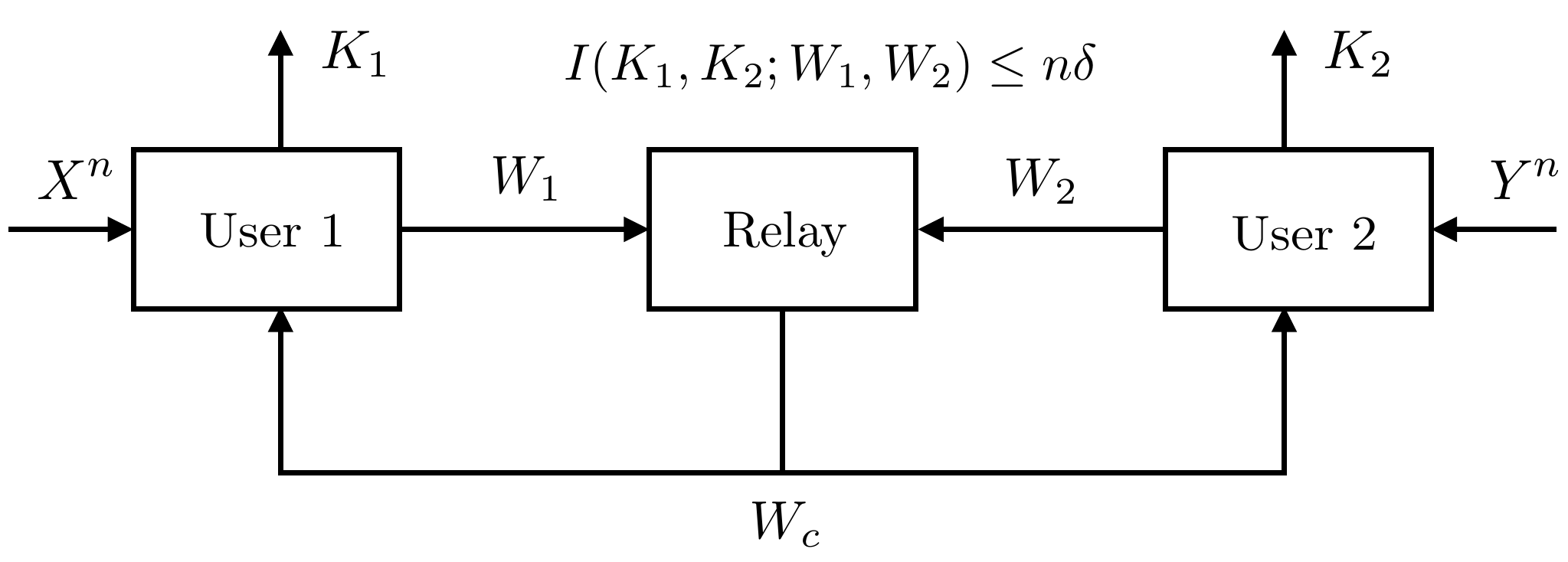}}
	\caption{Secret key generation through an untrusted relay.}
	\label{fig:model}
\end{figure}
From a network topology viewpoint, our problem is also related to two-way source coding through a relay \cite{ieTWSC} which has a different goal of reconstructing the other user's source sequence with the help of a relay. This include the special cases  
 \cite{kuMSCW08}, \cite{tgkLBWC} which consider the broadcast phase assuming that the relay knows both sources. 
Another view of establishing secret key agreement is through secure message transmission. Related works involving  the untrusted relay include, e.g., \cite{hyCWUR10,hmsSCVU13,pbSSCO11,rsejWSMU15}.

The rest of the paper is organized as follows. In Section II we consider the untrusted relay setting and present inner and outer bounds to the communication-key rate region. A Gaussian example is also discussed. Section III considers an extension where two users observe some common component and can therefore utilize it for secret key generation. Lastly, in Section IV, we consider the trusted relay setting with a public broadcast link. 
All notations follow standard ones in \cite{ekNIT11}.

\section{Secret Key Generation through an Untrusted Relay}

\subsection{Problem Formulation}\label{sec:problem_setting}
Let us consider secret key generation through a relay as depicted in Fig.~\ref{fig:model}. Source alphabets $\mathcal{X}, \mathcal{Y}$ are finite sets. Let $(X^n,Y^n)$ be $n$-length sequences which have i.i.d. components distributed according to some fixed joint distribution $P_{X,Y}$.

The  rate-limited descriptions $W_1$ and $W_2$ are generated based on $X^n$ and $Y^n$, respectively. The relay, after receiving $(W_1,W_2)$, generates another rate-limited description $W_c$ and broadcasts it to both users. User $1$ generates the key $K_1$ based on $X^n$ and $W_c$, while User $2$ generates the key $K_2$ based on $Y^n$ and $W_c$.  The goal is to establish key agreement between two users, i.e., $K_1=K_2$, with the highest key rate  while preserving privacy of the key by limiting the key leakage rate at the relay $\frac{1}{n}I(K_1,K_2;W_1,W_2)$ to a negligible level.  

We are interested in characterizing the optimal tradeoff among the communication rates of different  rate-limited links  and the resulting key rate.
\begin{definition}\label{def:code_mFAP}
	A $(|\mathcal{W}_1^{(n)}|,|\mathcal{W}_2^{(n)}|,|\mathcal{W}_c^{(n)}|,|\mathcal{K}^{(n)}|,n)$-code for secret key generation through a relay consists of
	\begin{itemize}
		\item an encoder  $f_1^{(n)}: \mathcal{X}^{n} \rightarrow \mathcal{W}_1^{(n)}$,
		\item an encoder  $f_2^{(n)}: \mathcal{Y}^{n} \rightarrow \mathcal{W}_2^{(n)}$,
		\item a relay mapping $f_r^{(n)}: \mathcal{W}_1^{(n)} \times \mathcal{W}_2^{(n)} \rightarrow \mathcal{W}_c^{(n)}$,
		\item a decoder  $g_1^{(n)}: \mathcal{W}_c^{(n)} \times \mathcal{X}^{n} \rightarrow\mathcal{K}^{(n)}$,
		\item a decoder  $g_2^{(n)}: \mathcal{W}_c^{(n)} \times \mathcal{Y}^{n} \rightarrow\mathcal{K}^{(n)}$.
		\hfill $\lozenge$
	\end{itemize}
\end{definition}

\begin{definition}  A rate tuple $(R_1,R_2,R_c,R_k) \in \mathbb{R}^{4}_{+}$ is said to be \emph{achievable} if, for any $\delta>0$ there exists a sequence of $(|\mathcal{W}_1^{(n)}|,|\mathcal{W}_2^{(n)}|,|\mathcal{W}_c^{(n)}|,|\mathcal{K}^{(n)}|,n)$-codes such that, for all sufficiently large $n$,  
	\begin{align}
	\mathrm{Pr}(K_1 \neq K_2) &\leq \delta\\
	\frac{1}{n}\log\big|\mathcal{W}_i^{(n)}\big| &\leq R_i+\delta,\ i=1,2,c\\
	\frac{1}{n}H(K_1)  &\geq R_k-\delta,  \\
	\frac{1}{n}I(K_1,K_2;W_1,W_2) &\leq \delta.  \label{eq:def_leakage}
	\end{align}
	The \emph{communication-key rate} region $\mathcal{R}$ is defined as the closure of the set of all achievable rate tuples.\hfill $\lozenge$
\end{definition}

\subsection{Results}
We provide inner and outer bounds for the communication-key rate region below. 
\begin{theorem}[Inner Bound]\label{theorem:inner}
	An inner bound $\mathcal{R}_{in}$ to the communication-key rate region  is given as the convex hull of a set of all tuples $(R_1,R_2,R_c,R_k) \in \mathbb{R}_{+}^4$ satisfying 
	\begin{align}
	R_1 &\geq I(X;U_1|Y) \label{eq:rate_R1}\\
	R_2 &\geq I(Y;U_2|X) \label{eq:rate_R2}\\
	R_c &\geq \max\{I(X;U_1|Y),I(Y;U_2|X)\} \label{eq:rate_Rc}\\
	R_k &\leq I(Y;U_1)+I(X;U_2)-I(U_1;U_2)\label{eq:rate_Rk1}\\
	&=I(X,Y;U_1,U_2)-I(X;U_1|Y)-I(Y;U_2|X),\label{eq:rate_Rk2}
	\end{align}
	for some $P_{X,Y}$$P_{U_1|X}P_{U_2|Y}$ with $|\mathcal{U}_1|\leq |\mathcal{X}|+1$ and $|\mathcal{U}_2|\leq |\mathcal{Y}|+1$.
\end{theorem}

\begin{IEEEproof}
	The proof idea is based on  an achievable scheme which combines binning, network coding, and key aggregation techniques. Each user compresses its source sequence using the Wyner-Ziv coding \cite{wzTRDF76} while treating the other source as side information. The relay simply combines two bin indices using index splitting and network coding techniques and broadcasts it to the two users. Each user, given the information from the relay and the observed source sequence, decodes the codeword chosen by the other user. The secret key is then formed as an aggregation of two partial codeword indices, one from his/her own codeword and another from the decoded codeword. Details of the proof are given in Appendix \ref{app:proof_inner}.
\end{IEEEproof}
\begin{remark}
	\begin{itemize}
		\item[1)] Since each user knows its chosen codeword, the relay only needs to send a modulo sum of the partial indices  for successful decoding at the users. This simple network coding technique at the relay helps to reduce the required rate  $R_c$. 
		\item[2)] The key generation process is done by concatenating two partial codeword indices to form a secret key. This process can provide a high total key rate and is justified by the ``independence" property of the two partial keys as shown in Lemma \ref{lemma:key_independence} in the proof of Theorem \ref{theorem:inner}. 
		\item[3)] The  key rate expression in \eqref{eq:rate_Rk1} appears to be less than the sum of key rates achievable from the one-way communication scheme, i.e., $I(Y;U_1)+I(X;U_2)$. Intuitively, a reduction $I(U_1;U_2)$ on the key rate  is to prevent the relay from learning the key completely by decoding the codewords from a given  $(W_1,W_2)$. 
		\item[4)] 
		Although our problem considers a different communication protocol involving a relay, 
		the key rate expression in \eqref{eq:rate_Rk2} resembles in some sense  the secret key capacity in \cite{cnSCFM04}, i.e., the joint entropy (rate for omniscience at some public terminal) minus the  rate needed for ``communication for omniscience" at users. In our case, due to the communication constraints, omniscience of the sources is replaced by that of the codewords $(U_1^n,U_2^n)$.
	\end{itemize} 
\end{remark}
\begin{remark}
	When there are no constraints on $R_1$, $R_2$, and $R_c$, the achievable key rate in Theorem~\ref{theorem:inner} reduces to $I(X;Y)$. This is in fact the secret key capacity for the classical two-user setting without communication constraint \cite{acCRIT93}, \cite{mSKAP93}. It can be obtained by setting $U_1=X$ and $U_2=Y$ in Theorem \ref{theorem:inner}.
\end{remark}
\begin{theorem}[Outer Bound]\label{theorem:outer}
	An outer bound $\mathcal{R}_{out}$  to the communication-key rate region is given as a set of all tuples $(R_1,R_2,R_c,R_k) \in \mathbb{R}_{+}^4$ satisfying \eqref{eq:rate_R1}-\eqref{eq:rate_Rc}, and
	\begin{align}
	R_k &\leq I(X,Y;U_1,U_2)-I(X;U_1|Y)-I(Y;U_2|X),
	\end{align}
	for some $P_{X,Y}$$P_{U_1,U_2|X,Y}$.
\end{theorem}
\begin{IEEEproof}
	 The proof follows from standard properties of the entropy function with Fano's inequality 
	  $H(K_1|W_c,Y^n) \leq n\epsilon_n$ and $H(K_2|W_c,X^n) \leq n\epsilon_n$ 
	 and the key leakage constraint \eqref{eq:def_leakage}. The details of the proof are given in Appendix~\ref{app:proof_outer}.
\end{IEEEproof}
\begin{remark}
The rate expressions in Theorems  \ref{theorem:inner} and \ref{theorem:outer}  are the same and the only difference is in the set of probability distributions.  We note that from the problem formulation in Section \ref{sec:problem_setting}, $K_1$ and $K_2$ are generated based on both sources and $W_c$. The dependence of $W_c$ corresponds to utilizing feedback information which  creates difficulty in deriving the tight bound. In particular, the auxiliary random variables to be defined in the converse proof often contain the key variables which are dependent of $W_c$ and both sources. 
\end{remark}

\subsection{Example and Discussion}
We consider a Gaussian example of the inner bound given in Theorem~\ref{theorem:inner} and compare it to the results obtained from a simple one-way setting where the relay only forwards information in one direction. 
Let $\begin{bmatrix}
X \\ 
Y
\end{bmatrix}  \sim \mathcal{N}\Bigg(\begin{bmatrix}
0 \\ 
0
\end{bmatrix} ,\begin{bmatrix}
1 & \rho \\ 
\rho & 1
\end{bmatrix}\Bigg)$, where $\rho \in [0,1]$. We can 
choose $U_1 = X+Q_1$, and $U_2 = Y+Q_2$, where $Q_1 \sim \mathcal{N}(0,N_{Q_1})$ and $Q_2 \sim \mathcal{N}(0,N_{Q_2})$ are independent of each other and of $(X,Y)$.  It can be shown that Theorem  \ref{theorem:inner} reduces to the set of $(R_1,R_2,R_c,R_k)$ such that 
\begin{align*}
R_i &\geq \frac{1}{2}\log\Big(\frac{1+N_{Q_i}-\rho^2}{N_{Q_i}}\Big),\ i=1,2\\
R_c &\geq \max\Big\{ \frac{1}{2}\log\Big(\frac{1+N_{Q_1}-\rho^2}{N_{Q_1}}\Big), \frac{1}{2}\log\Big(\frac{1+N_{Q_2}-\rho^2}{N_{Q_2}}\Big)\Big\}\\
R_k &\leq  \frac{1}{2}\log\bigg(\frac{(1+N_{Q_1})(1+N_{Q_2})-\rho^2}{(1+N_{Q_1}-\rho^2)(1+N_{Q_2}-\rho^2)}\bigg),
\end{align*}
for some $N_{Q_1}, N_{Q_2} \in \mathbb{R}^+$.

Furthermore, by setting $N_{Q_i} = \frac{1-\rho^2}{2^{2\min\{R_i,R_c\}}-1}$ for $i=1,2$, 
we have that 
\begin{equation}\label{eq:max-key-rate-Gaussian}
R_k \leq \frac{1}{2}\log\bigg(\frac{1-\rho^2 2^{-2(\min\{R_1,R_c\}+\min\{R_2,R_c\})}}{1-\rho^2}\bigg).
\end{equation}
\begin{figure}[]
	\centerline{\includegraphics[width=19cm]{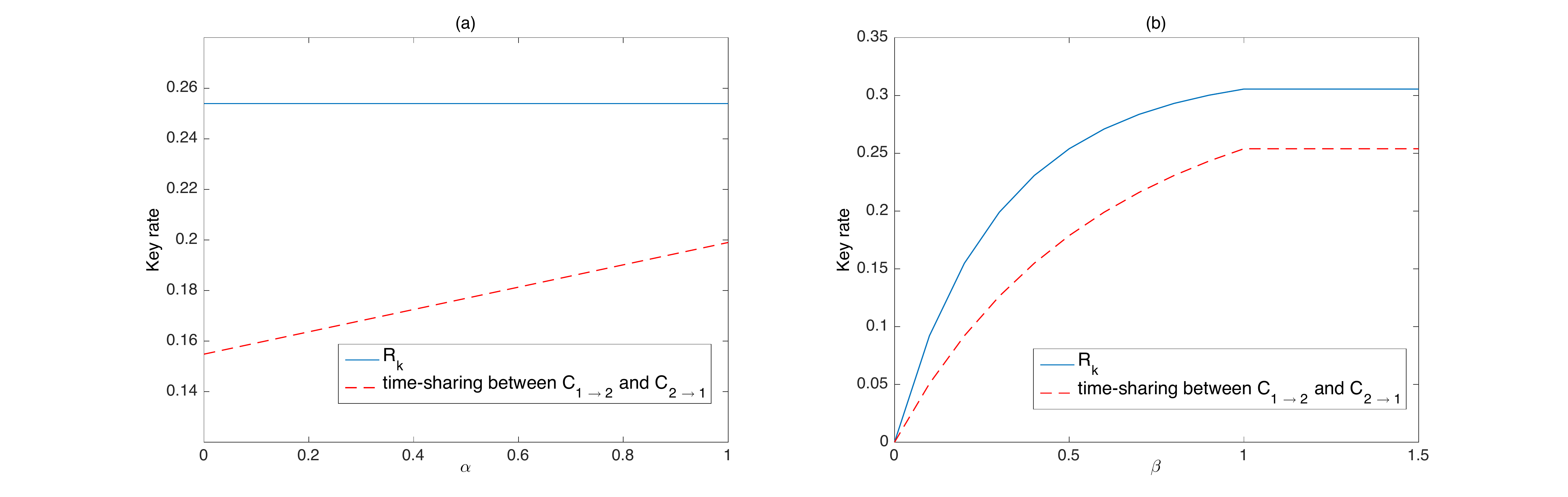}}
	\caption{Comparison between achievable secret key rate $R_k$ in \eqref{eq:max-key-rate-Gaussian} and a convex combination of $C_{1\rightarrow 2}$ and $C_{2\rightarrow 1}$ in \eqref{eq:oneway_key_capacity}, i.e., $C^* = \alpha C_{1 \rightarrow 2} + (1-\alpha)C_{2 \rightarrow 1}$. In (a), we assume that $R_1=0.6$, $R_2=0.4$, $R_c =1$,  and $\rho =0.6$. In (b), we assume that $R_1=R_2= \beta R_c$ where $R_c =1$ and $\rho =0.6$, and here we see that the key rate is saturated at $\beta \geq 1$ as the common  rate $R_c$ becomes a bottleneck.}
	\label{fig:plot}
\end{figure}
Note that for the one-way communication system (from $i$ to $j$), we can derive the expression for the secret key capacity in terms of communication rates (similarly as in \cite{woSKAF10}), i.e.,
\begin{align}
C_{i\rightarrow j} &=  \frac{1}{2}\log\bigg(\frac{1-\rho^2 2^{-2\min\{R_i,R_c\}}}{1-\rho^2}\bigg),\label{eq:oneway_key_capacity}
\end{align}
for $i,j \in \{1,2\}$, $i \neq j$.

We see that for given $(R_1,R_2,R_c)$, the maximum achievable key rate in \eqref{eq:max-key-rate-Gaussian} 
can be strictly larger than the convex combination of the one-way secret key capacities (see, e.g., Fig. \ref{fig:plot} (a) and (b) for simple illustration). This suggests that there is some benefit in utilizing the communications through a relay in establishing the secret key agreement as compared to the classical one-way scheme  in general.

\section{Sources With Common Components}
Next we consider an extension where we assume that the source sequences consist of a common part $Z^n$. Without loss of generality, we  assume that User 1 observes $(X^n,Z^n)$ and User 2 observes $(Y^n,Z^n)$, where $(X^n,Y^n,Z^n)$ are i.i.d. according to $P_{X,Y,Z}$. 
The problem formulation for this case  is essentially the same as in the previous section where we replace $X^n$ by $(X^n,Z^n)$ and $Y^n$ by $(Y^n,Z^n)$.  

It is interesting to see how the users can utilize the common part of the sources for secret key generation. One simple strategy is to let the users exclusively use the common source $Z^n$ to generate the secret key without sending any information to the relay. This method  achieves the secret key rate of $H(Z)$ without leaking any information. We show that when the sources are \emph{conditionally independent}, i.e., $X-Z-Y$ forms a Markov chain, this strategy is in fact optimal. However, in general, the users could benefit 
if they communicate through a relay. In this section, we provide an inner bound to the communication-key rate region. Our achievable scheme is based on the idea of utilizing the common source to generate a partial key which can be combined with the key  generated from the agreement through the relay. The scheme is a direct extension of that in the previous section. Consequently, the inner bound here recovers Theorem~\ref{theorem:inner} when $Z$ is constant. 

\begin{theorem}[Inner Bound]\label{theorem:inner2}
	An inner bound $\mathcal{R}_{in,common}$ to the communication-key rate region  is given as the convex hull of a set of all tuples $(R_1,R_2,R_c,R_k) \in \mathbb{R}_{+}^4$ satisfying 
	\begin{align}
	R_1 &\geq I(X;U_1|Y,Z) \\
	R_2 &\geq I(Y;U_2|X,Z) \\
	R_c &\geq \max\{I(X;U_1|Y,Z),I(Y;U_2|X,Z)\} \\
	R_k &\leq I(Y,Z;U_1)+I(X,Z;U_2)-I(U_1;U_2) +H(Z|U_1,U_2)
	\end{align}
	 for some $P_{X,Y,Z}$$P_{U_1|X,Z}P_{U_2|Y,Z}$ with $|\mathcal{U}_1|\leq |\mathcal{X}||\mathcal{Z}|+2$ and $|\mathcal{U}_2|\leq |\mathcal{Y}||\mathcal{Z}|+2$.
\end{theorem}

\begin{IEEEproof}
	The proof idea is based on an extension of the achievable scheme used to prove Theorem \ref{theorem:inner}. We utilize the common source $Z^n$ to generate an additional part of secret key at rate close to  $H(Z|U_1,U_2)$ and combine it with the key  generated as in the previous scheme. The encoding, relay mapping, and decoding processes are similar as before, except that each user operates on the ``super-sources" $(X^n,Z^n)$ and $(Y^n,Z^n)$ instead. 
		In Appendix \ref{app:proof_inner2}, we provide the proof of achievable key rate and analysis of the key leakage rate. 
\end{IEEEproof}

\begin{corollary}
	If the sources are conditionally independent, i.e., $X-Z-Y$, we have that the secret key capacity is given by $H(Z)$. This result is very intuitive since the users observe the common source $Z^n$, and given $Z^n$, the sources $X^n$ and $Y^n$ are independent. Therefore, $X^n$ and $Y^n$ do not contribute to the key generation process. We can simply generate the secret key by using only the common source, i.e., hashing (binning) the sequence $Z^n$.  
\end{corollary}

\section{Trusted Relay with Public Broadcast Transmission}\label{sec:trusted}
\begin{figure}[]
	\centerline{\includegraphics[width=9.7cm]{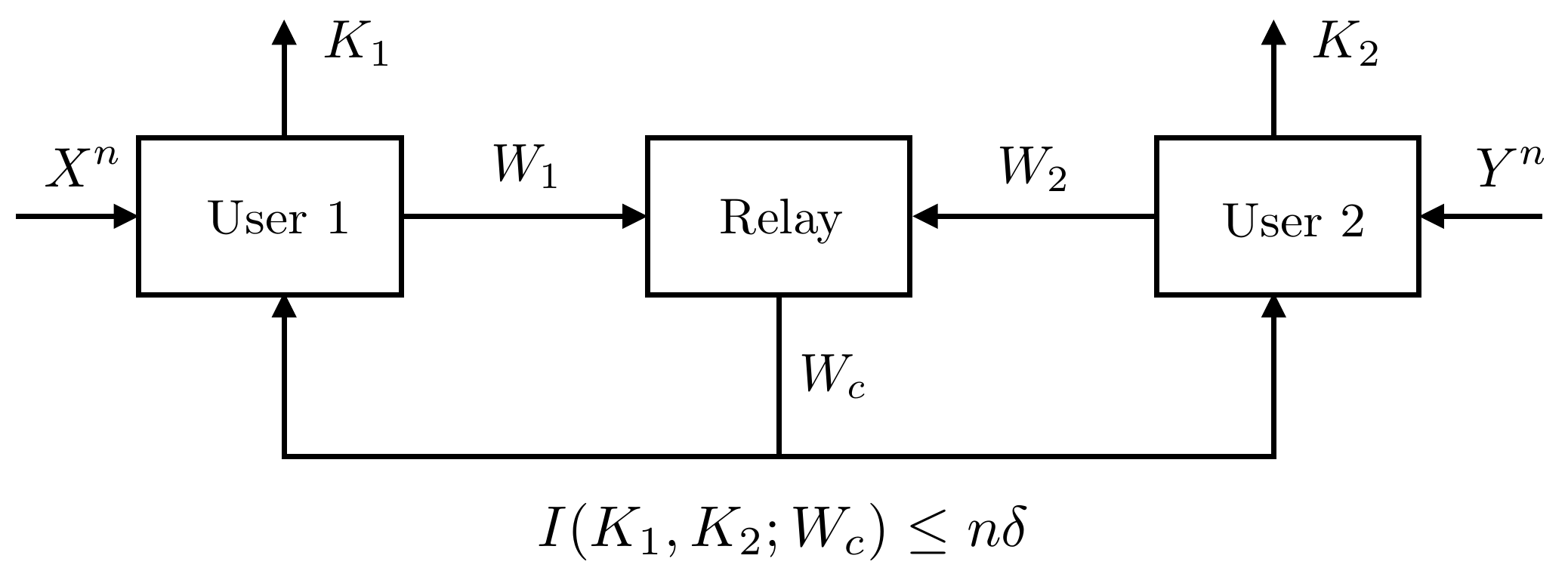}}
	\caption{Secret key generation through a trusted relay.}
	\label{fig:model2}
\end{figure}

In this section, we consider a new setting of secret key generation through a relay where the relay is trustworthy but its transmission to the users can be eavesdropped upon by an external passive eavesdropper, as depicted in Fig. \ref{fig:model2}. 
The problem formulation remains the same as in the previous one in Section \ref{sec:problem_setting}, except that the constraint on  key leakage rate becomes $\frac{1}{n}I(K_1,K_2;W_c) \leq \delta$. This  constraint can be seen as  a weaker version of \eqref{eq:def_leakage} since it can be implied by \eqref{eq:def_leakage}.

We note that, by moving the key leakage constraint from the relay,  the nature of the problem changes quite drastically.  For instance, it is now possible for the relay to decode some codewords chosen at the users without violating the key leakage constraint. 

\subsection{Result}\label{sec:trusted_result}
\begin{figure}[]
	\centerline{\includegraphics[width=9.5cm]{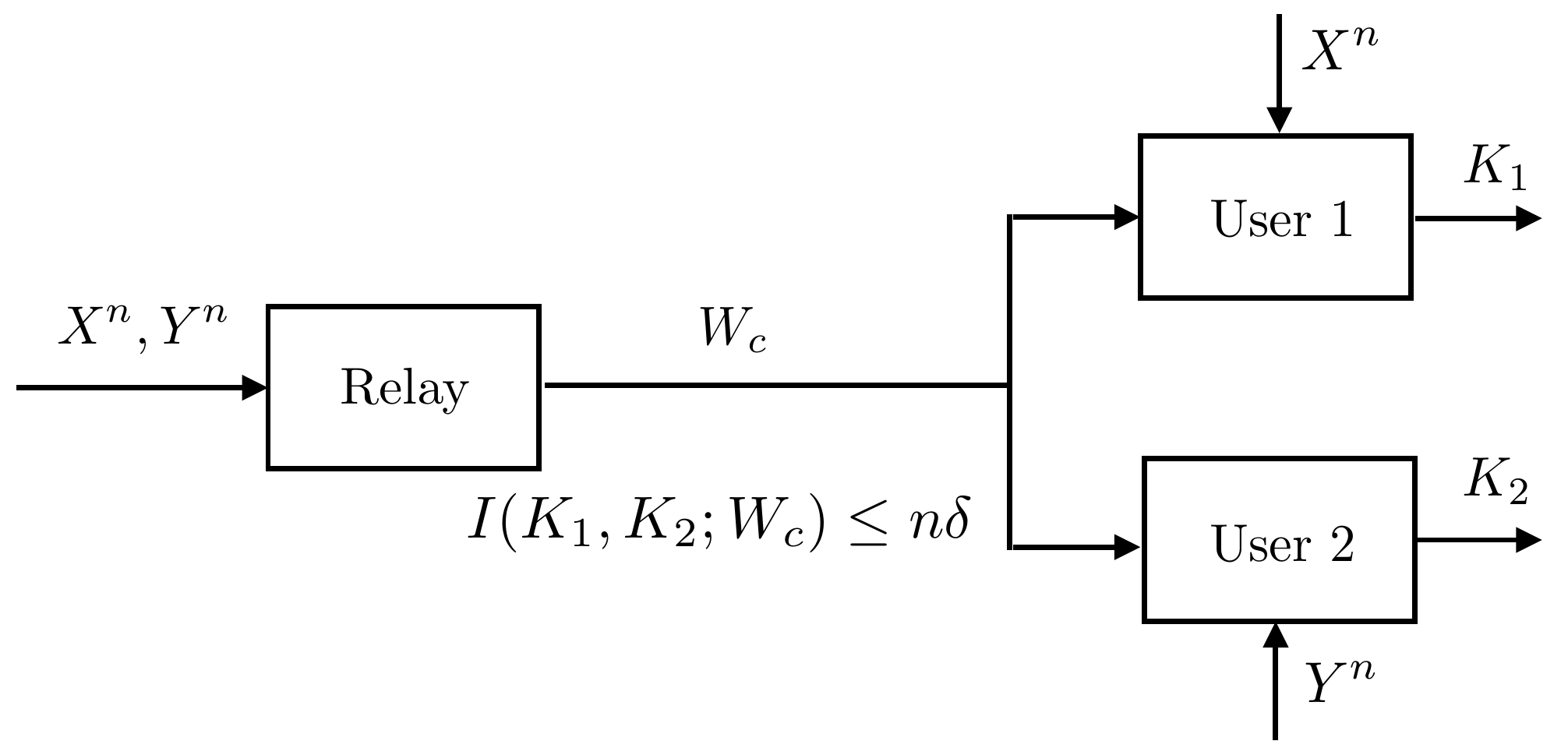}}
	\caption{Secret key generation when sources are available at relay.}
	\label{fig:model3}
\end{figure}

In the following, we present the communication-key rate region for the special case, depicted in Fig.~\ref{fig:model3}, where the uplink communications are over links with unlimited capacities, i.e.,  $(X^n,Y^n)$ is available at the relay. 

\begin{theorem}\label{theorem:region_special_case}
	The communication-key rate region for the setting in Fig. \ref{fig:model3} is  the set of all $(R_c,R_k) \in \mathbb{R}_+^2$  such that 
	\begin{align}
	R_c &\geq \max\{I(X;V|Y),I(Y;V|X)\}\\
	R_k &\leq \min\{I(X;V),I(Y;V)\}
	\end{align}
	for some $P_{X,Y}P_{V|X,Y}$ with $|\mathcal{V}| \leq |\mathcal{X}||\mathcal{Y}|+2$.
\end{theorem}
\begin{IEEEproof}
	The achievability is based on the Wyner-Ziv coding with respect to side information at the users. The bin index of the codeword is sent over the rate-limited link and the codeword index is selected as a secret key. With $R_c \geq \max\{I(X;V|Y),I(Y;V|X)\}$, we ensure that both users can decode the codeword and therefore  agree on the common secret key. The converse  follows from the key leakage rate constraint, Fano's inequality, and some standard properties of the entropy function. 
	Proof details are given  in Appendix \ref{app:proof_trusted_relay}.
\end{IEEEproof}

\subsection{Comparison to the Previous Scheme}
For the general setting in Fig. \ref{fig:model2}, in view of minimizing the rates $R_1$ and $R_2$, one may devise a coding scheme where the first phase of transmission (from the users to the relay) is based on multi-terminal lossy source coding \cite{bMSC77}. The relay, after decoding, can choose another codeword to communicate to users to establish the secret key generation. If the rates $R_1$ and $R_2$ are sufficiently high, then the first phase transmission can support lossless reconstruction, i.e., the scheme reduces to the Slepian-Wolf coding \cite{swNCOC73}, and the relay can now operate directly on the sources as considered in the special case in Section \ref{sec:trusted_result}. 

It is however unclear whether this type of scheme where the relay first decodes some codewords is a good scheme that can achieve the optimal tradeoff in general. 
For comparison, we discuss the  scheme used to prove Theorem~\ref{theorem:inner} where the relay does not decode any codeword, but rather employs a network coding technique to forward information to the users. It can be shown that the inner bound in Theorem~\ref{theorem:inner} also holds for the trusted relay setting since  $I(K_1,K_2;W_1,W_2)=I(K_1,K_2;W_1,W_2,W_c)\geq I(K_1,K_2;W_c) $.  However, when specializing to the case of unlimited uplink rates, the inner bound in Theorem~\ref{theorem:inner} reduces to the set of $(R_c,R_k)$ satisfying
	\begin{align*}
	R_c &\geq 
	\max\{I(X;U_1,U_2|Y),I(Y;U_1,U_2|X)\},\\
	R_k &\leq  I(X;U_1,U_2)-I(X;U_1,U_2|Y)\\
	& = I(Y;U_1,U_2)-I(Y;U_1,U_2|X),
	\end{align*}
	for some $P_{X,Y}$$P_{U_1|X}P_{U_2|Y}$.

We see that the region in Theorem \ref{theorem:region_special_case} is generally larger than the one above. 
This suggests that for the trusted relay setting, the scheme where the relay first decodes some codewords can perform better in general.

	\appendices			

\section{Proof of Theorem \ref{theorem:inner}}\label{app:proof_inner}
The proof is based on a random coding argument where we follow the definition and properties of joint typicality in \cite{ekNIT11}.

\textit{Codebook generation}: Fix $P_{U_1|X}P_{U_2|Y}$. 
\begin{itemize}
	\item Randomly and independently generate codewords $u_1^n(w_{1a},w_{1b},w_{1k},w')$ each according to $P_{U_1}$, where $w_{1a} \in [1:2^{n(I(X;U_1|Y)-R_b+2\delta_{\epsilon})}]$, $w_{1b} \in [1:2^{nR_b}]$, $w_{1k}\in[1:2^{nR_{k1}}]$, and $w' \in [1:2^{n(I(Y;U_1)-R_{k1}-\delta_{\epsilon})}]$.
	\item Randomly and independently generate codewords $u_2^n(w_{2a},w_{2b},w_{2k},w'')$ each according to $P_{U_2}$, where $w_{2a} \in [1:2^{n(I(Y;U_2|X)-R_b+2\delta_{\epsilon})}]$, $w_{2b} \in [1:2^{nR_b}]$, $w_{2k}\in[1:2^{nR_{k2}}]$, and $w'' \in [1:2^{n(I(X;U_2)-R_{k2}-\delta_{\epsilon})}]$.
	\item Let $R_b = \min\{I(X;U_1|Y),I(Y;U_2|X)\}$ and $R_{k1}+R_{k2} = I(Y;U_1)+I(X;U_2)-I(U_1;U_2)-\delta_{\epsilon} >0$.
	\item The codebook is revealed to both users and the relay.
\end{itemize}

\textit{Encoding:}
\begin{itemize}
	\item User 1: Given $x^n$, it looks for a jointly typical $u_1^n$. From the covering lemma \cite{ekNIT11}, with high probability, there exists at least one such codeword. If there are more than one, it selects the one with the smallest indices. Then it sends the indices $(w_{1a},w_{1b})$ to the relay.
	\item Encoding for User 2 follows similarly as that for User~1, but with $y^n$ and $u_2^n$ instead of $x^n$ and $u_1^n$.  Finally, it sends the indices $(w_{2a},w_{2b})$ to the relay.
	\item By the Markov lemma \cite{ekNIT11}, $(x^n,y^n,u_1^n,u_2^n)$ are jointly typical with high probability.
\end{itemize}

\textit{Relay mapping}: Given $(w_{1a},w_{1b},w_{2a},w_{2b})$, the relay broadcasts  $w_c = (w_{1a},w_{1b}\oplus w_{2b},w_{2a})$ back to the users, where $\oplus$ denotes the addition over $2^{nR_b}$ field.

\textit{Decoding:} 
\begin{itemize}
	\item User 1: Given $w_c$ and $(x^n,u_1^n(w_{1a},w_{1b},w_{1k},w'))$, it can decrypt $w_{2b}$. Then it looks for a unique $(\hat{w}_{2k},\hat{w}'')$ such that $u_2^n$ is jointly typical with $(x^n,u_1^n)$. From the packing lemma \cite{ekNIT11}, with high probability, it finds such a pair and it is the correct pair selected at User 2. User~1 then generates the key $k_1=(w_{1k},\hat{w}_{2k})$.
	\item Similarly as  User 1, User 2 finds a unique $(\hat{w}_{1k},\hat{w}')$ and generates the key as $k_2=(\hat{w}_{1k},w_{2k})$. With high probability, $k_1=k_2=(w_{1k},w_{2k})$.
\end{itemize} 
Let $U_1^n(W_{1a},W_{1b},W_{1k},W')$ and $U_2^n(W_{2a},W_{2b},W_{2k},W'')$ be the chosen codewords in the encoding process. From the LLN, we have that  sequences $(X^n,Y^n,U_1^n(W_{1a},W_{1b},W_{1k},W'),U_2^n(W_{2a},W_{2b},W_{2k},W''))$ are jointly typical with high probability.

\textit{Analysis of key leakage}: Let $W_1=(W_{1a},W_{1b})$ and $W_2=(W_{2a},W_{2b})$. The key leakage averaged over all randomly chosen codebooks can be bounded as follows:  
\begin{align}
&I(K_1,K_2;W_1,W_2)=I(W_{1k},W_{2k};W_1,W_2)\nonumber\\
&\leq H(W_{1k},W_{2k})-I(W_{1k},W_{2k};X^n,Y^n|W_1,W_2)\nonumber\\
&\leq H(W_{1k},W_{2k})-H(X^n,Y^n)+H(W_1,W_2) +H(X^n,Y^n|W_1,W_2,W_{1k},W_{2k},W',W'')\nonumber\\
&\qquad+H(W',W''|W_1,W_2,W_{1k},W_{2k})\nonumber\\
&\overset{(a)}{\leq} n(I(Y;U_1)+I(X;U_2)-I(U_1;U_2)-\delta_{\epsilon}) - nH(X,Y) + n(I(X;U_1|Y)+2\delta_{\epsilon}) \nonumber\\
&\qquad + n(I(Y;U_2|X)+2\delta_{\epsilon})  + n(H(X,Y|U_1,U_2)+\delta_{\epsilon})  +n\epsilon_{n} \overset{(b)}{\leq} n\delta_{\epsilon}',\nonumber
\end{align}
where  $(a)$ follows from the codebook generation, from the bound $H(X^n,Y^n|W_1,W_2,W_{1k},W_{2k},W',W'') \leq n(H(X,Y|U_1,U_2)$$+\delta_{\epsilon})$ which follows from properties of jointly typical sequences (the proof is given in Appendix \ref{app:proof_bound_conditional_entropy}), and from Fano's inequality $H(W',W''|W_1,W_2,W_{1k},W_{2k}) \leq n\epsilon_{n}$ which holds since given $(W_1,W_2,W_{1k},W_{2k})$, the codewords $(U_1^n,U_2^n)$ and thus $(W',W'')$ can be decoded successfully with high probability (the mutual packing lemma \cite{ekNIT11}). Lastly, $(b)$ follows from  $U_1-X-Y-U_2$.

Before proceeding to the analysis of an achievable key rate, we provide a lemma which states the ``independence" property of the two index parts $(W_{1k},W_{2k})$ that form a secret key.
\begin{lemma}\label{lemma:key_independence}
	With the achievable scheme and codebook generated as described above, we have that $I(W_{1k};W_{2k}) \leq n\delta_{\epsilon}$.
\end{lemma}

The proof of the lemma is given as follows:
\begin{align*}
&	I(W_{1k};W_{2k}) 
\leq H(W_{1k})-I(W_{1k};X^n|W_{2k})\\
&= H(W_{1k})-H(X^n)-H(W_{2k}|X^n)+H(W_{2k}) +H(X^n|W_{1k},W_{2k},W_1,W_2,W',W'')\\ &\qquad + I(X^n;W_1,W_2,W',W''|W_{1k},W_{2k})\\
&\overset{(a)}{\leq} H(W_{1k})-H(X^n)-H(W_{2k}|X^n)+H(W_{2k}) +n(H(X|U_1,U_2)+\delta_{\epsilon})\\&\qquad + H(W_1,W_2,W',W''|W_{1k},W_{2k})-H(W_1,W_2,W',W''|W_{1k},W_{2k},X^n)\\
&\overset{(b)}{\leq}  H(W_{1k})-H(X^n)-I(W_{2k};Y^n|X^n)+H(W_{2k}) +n(H(X|U_1,U_2)+\delta_{\epsilon})+ H(W_1,W_2|W_{1k},W_{2k})\\ &\qquad +n\epsilon_{n} -I(W_2,W'';Y^n|W_{2k},X^n)\\
&\overset{(c)}{\leq}  H(W_{1k})+H(W_{2k})-H(X^n,Y^n)+n(H(X|U_1,U_2)+\delta_{\epsilon})+ H(W_1)+H(W_2) +n\epsilon_{n}\\ &\qquad+n(H(Y|U_2,X)+\delta_{\epsilon})\\
&\overset{(d)}{=} 	H(W_{1k})+H(W_{2k})-nI(X,Y;U_1,U_2)+ H(W_1) +H(W_2) +n\delta_{\epsilon}'\\
& \overset{(e)}{\leq} n\delta_{\epsilon}'',	 
\end{align*}
where $(a)$ follows from the bound on the conditional entropy term $H(X^n|W_{1k},W_{2k},W_1,W_2,W',W'') \leq n(H(X|U_1,U_2)+\delta_{\epsilon})$ (see, e.g., Appendix \ref{app:proof_bound_conditional_entropy}), $(b)$ follows from Fano's inequality $H(W',W''|W_1,W_2,W_{1k},W_{2k}) \leq n\epsilon_{n}$ and the fact that $(W_1,W_{1k},W')$ is a function of $X^n$, $(c)$ follows from the bound $H(Y^n|W_2,W_{2k},W'',X^n) \leq n(H(Y|U_2,X)+\delta_{\epsilon})$ (see, e.g., Appendix \ref{app:proof_bound_conditional_entropy}), $(d)$ follows from the Markov chain $U_1-(X,U_2)-Y$, and $(e)$ follows from the codebook generation.

Next we provide the analysis of achievable key rate.
\begin{align*}
&H(K_1) = H(W_{1k},W_{2k})
\overset{(a)}{\geq} H(W_{1k})+H(W_{2k})-n\delta_{\epsilon}\\
&\geq H(W_{1k},W_1,W')-H(W_1)-H(W') + H(W_{2k},W_2,W'')-H(W_2)-H(W'')-n\delta_{\epsilon}\\
&\overset{(b)}{\geq} H(U_1^n)-H(W_1)-H(W') + H(U_2^n)-H(W_2)-H(W'')-n\delta_{\epsilon}\\
&\overset{(c)}{\geq} n(I(X;U_1)+\delta_{\epsilon})-H(W_1)-H(W')  + n(I(Y;U_2)+\delta_{\epsilon})-H(W_2)-H(W'')-n\delta_{\epsilon}\\
&\overset{(d)}{\geq} n(I(Y;U_1)+I(X;U_2)-I(U_1;U_2)-\delta_{\epsilon}')
\geq n(R_k-\delta_{\epsilon}')
\end{align*}
if $R_k \leq I(Y;U_1)+I(X;U_2)-I(U_1;U_2)$, 
where $(a)$ follows from Lemma \ref{lemma:key_independence}, $(b)$ follows from the fact that given the codebook, $U_1^n$ is a function of $(W_{1k},W_1,W')$, and similarly for $U_2^n$, $(c)$ follows from the encoding processes where $\mathrm{Pr}(U_1^n=u_1^n) \leq 2^{-n(I(X;U_1)-\delta_{\epsilon})}$ and $\mathrm{Pr}(U_2^n=u_2^n) \leq 2^{-n(I(Y;U_2)-\delta_{\epsilon})}$, and $(d)$ follows from the codebook generation. The cardinality bound can be proved using the support lemma (see, e.g., \cite{ekNIT11}). 

				\section{Proof of Theorem \ref{theorem:outer}}\label{app:proof_outer}
				Let $U_{1,i} = (K_1,W_2,W_c,Y^{i-1})$ and $U_{2,i}=(K_2,W_1,W_c,X^{i-1})$. For any achievable rate tuple $(R_1,R_2,R_c,R_k)$, it follows that 
				\begin{align*}
				nR_1 &\geq H(W_1)\geq H(W_1|Y^n)-H(W_1,K_1|X^n,Y^n)\\
				&= H(W_1,K_1|Y^n)-H(K_1|W_1,Y^n) -H(W_1,K_1|X^n,Y^n)\\
				&\overset{(a)}{\geq}  I(W_1,K_1;X^n|Y^n)-n\epsilon_n\\ 
				&\overset{(b)}{=}H(X^n|Y^n) -H(X^n|W_1,W_2,W_c,K_1,Y^n) -n\epsilon_n\\
				&\overset{(c)}{\geq} \sum_{i=1}^n I(X_i;U_{1,i}|Y_i)-n\epsilon_n,
				\end{align*}
				where $(a)$ follows from the fact that $W_2$ is a function of $Y^n$ and that $W_c$ is a function of $(W_1,W_2)$. Then by Fano's inequality, we have $H(K_1|W_c,Y^n) \leq n\epsilon_n$, $(b)$ follows from the fact that $W_2$ is a function of $Y^n$ and that $W_c$ is a function of $(W_1,W_2)$, and $(c)$ follows from the definition of $U_{1,i}$.
				
				Similarly, 
				\begin{align*}
				nR_2 &\geq H(W_2)\geq H(W_2|X^n)-H(W_2,K_2|X^n,Y^n)\\
				&= H(W_2,K_2|X^n)-H(K_2|W_2,X^n) -H(W_2,K_2|X^n,Y^n)\\
				&\overset{(a)}{\geq}  I(W_2,K_2;Y^n|X^n)-n\epsilon_n\\ 
				&\overset{(b)}{=}H(Y^n|X^n) -H(Y^n|W_1,W_2,W_c,K_2,X^n) -n\epsilon_n\\
				&\overset{(c)}{\geq} \sum_{i=1}^n I(Y_i;U_{2,i}|X_i)-n\epsilon_n,
				\end{align*}
				where $(a)$ follows from the fact that $W_1$ is a function of $X^n$ and that $W_c$ is a function of $(W_1,W_2)$. Then by Fano's inequality, we have $H(K_2|W_c,X^n) \leq n\epsilon_n$, $(b)$ follows from the fact that $W_1$ is a function of $X^n$ and that $W_c$ is a function of $(W_1,W_2)$, and $(c)$ follows from the definition of $U_{2,i}$.
				
				Next,
				\begin{align*}
				nR_c &\geq H(W_c)\geq H(W_c|Y^n)-H(W_c,K_1|X^n,Y^n)\\
				&\overset{(a)}{\geq} I(W_c,K_1;X^n|Y^n)-n\epsilon_n\\
				&\overset{(b)}{=} H(X^n|Y^n)-H(X^n|W_c,K_1,W_2,Y^n)-n\epsilon_n\\
				&\overset{(c)}{\geq} \sum_{i=1}^n I(X_i;U_{1,i}|Y_i)-n\epsilon_n,
				\end{align*} 
				where $(a)$ follows from Fano's inequality $H(K_1|W_c,Y^n) \leq n\epsilon_n$, $(b)$ follows from the fact that $W_2$ is a function of $Y^n$, and $(c)$  follows from the definition of $U_{1,i}$.
				And similarly, 
				\begin{align*}
				nR_c &\geq H(W_c)\geq H(W_c|X^n)-H(W_c,K_2|X^n,Y^n)\\
				&\overset{(a)}{\geq} I(W_c,K_2;Y^n|X^n)-n\epsilon_n\\
				&\overset{(b)}{=} H(Y^n|X^n)-H(Y^n|W_c,K_2,W_1,X^n)-n\epsilon_n\\
				&\overset{(c)}{\geq} \sum_{i=1}^n I(Y_i;U_{2,i}|X_i)-n\epsilon_n,
				\end{align*} 
				where $(a)$ follows from Fano's inequality $H(K_2|W_c,X^n) \leq n\epsilon_n$, $(b)$ follows from the fact that $W_1$ is a function of $X^n$, and $(c)$  follows from the definition of $U_{2,i}$.
				
				Lastly,
				\begin{align*}
				nR_k &\leq H(K_1)\leq H(K_1,K_2)
				= H(K_1,K_2,W_1,W_2) - H(W_1,W_2|K_1,K_2)\\
				&\leq I(K_1,K_2,W_1,W_2;X^n,Y^n) - H(W_1,W_2|K_1,K_2)\\
				&\overset{(a)}{\leq} I(K_1,K_2,W_1,W_2;X^n,Y^n) - H(W_1,W_2) + n\delta_n\\
				&\overset{(b)}{=}  \sum_{i=1}^n I(X_i,Y_i;U_{1,i},U_{2,1}) - H(W_1)-H(W_2) + I(W_1;W_2)+ n\delta_n\\
				&\leq \sum_{i=1}^n I(X_i,Y_i;U_{1,i},U_{2,1}) - H(W_1|Y^n)-I(W_1;Y^n)+H(W_1,K_1|X^n,Y^n)-H(W_2|X^n) \\&\qquad  -I(W_2;X^n)+H(W_2,K_2|X^n,Y^n) + I(W_1;W_2)+ n\delta_n\\
				&\overset{(c)}{\leq} \sum_{i=1}^n I(X_i,Y_i;U_{1,i},U_{2,1}) -  I(X_i;U_{1,i}|Y_i)-I(Y_i;U_{2,i}|X_i) -I(W_1;Y^n)-I(W_2;X^n) + I(W_1;W_2)+ n\delta_n\\
				&\overset{(d)}{\leq} \sum_{i=1}^n I(X_i,Y_i;U_{1,i},U_{2,1}) -  I(X_i;U_{1,i}|Y_i)-I(Y_i;U_{2,i}|X_i) + n\delta_n,
				\end{align*}
				where $(a)$ follows from the key leakage constraint, $(b)$ follows from the fact that $W_c$ is a function of $(W_1,W_2)$ and the definitions of $U_{1,i}$ and $U_{2,i}$, $(c)$ follows from the bounds on $R_1$ and $R_2$ above, and $(d)$ follows from the bound $-I(W_1;Y^n)-I(W_2;X^n) + I(W_1;W_2) \leq 0$ which holds due to the Markov chain $W_1-Y^n-W_2$.
				
				The proof ends by standard steps of introducing a time-sharing random variable and letting $n \rightarrow \infty$.

				\section{Proof of Achievable Key Rate and Analysis of Key Leakage Rate in Theorem \ref{theorem:inner2}}\label{app:proof_inner2}
				In the codebook generation, generating  codewords $u_1^n$ and $u_2^n$ with appropriate sizes similarly as in the proof of Theorem~\ref{theorem:inner}, i.e., 	
				\begin{itemize}
					\item Randomly and independently generate codewords $u_1^n(w_{1a},w_{1b},w_{1k},w')$ each according to $P_{U_1}$, where $w_{1a} \in [1:2^{n(I(X;U_1|Y,Z)-R_b+2\delta_{\epsilon})}]$, $w_{1b} \in [1:2^{nR_b}]$, $w_{1k}\in[1:2^{nR_{k1}}]$, and $w' \in [1:2^{n(I(Y,Z;U_1)-R_{k1}-\delta_{\epsilon})}]$.
					\item Randomly and independently generate codewords $u_2^n(w_{2a},w_{2b},w_{2k},w'')$ each according to $P_{U_2}$, where $w_{2a} \in [1:2^{n(I(Y;U_2|X,Z)-R_b+2\delta_{\epsilon})}]$, $w_{2b} \in [1:2^{nR_b}]$, $w_{2k}\in[1:2^{nR_{k2}}]$, and $w'' \in [1:2^{n(I(X,Z;U_2)-R_{k2}-\delta_{\epsilon})}]$.
					\item Let $R_b = \min\{I(X;U_1|Y,Z),I(Y;U_2|X,Z)\}$ and $R_{k1}+R_{k2} = I(Y,Z;U_1)+I(X,Z;U_2)-I(U_1;U_2)-\delta_{\epsilon} >0$.
				\end{itemize}
				
				Apart from that, we also partition the set $\mathcal{Z}^n$ by distributing sequences $z^n \in \mathcal{Z}^n$ uniformly at random into $2^{nR_{k,z}}$ equal-sized bins, where $R_{k,z}=H(Z|U_1,U_2)-2\delta_{\epsilon}$. Each user, knowing $z^n$, can find the corresponding bin index containing $z^n$ and set it to be a partial key $K_{z}$.  Following the similar coding scheme for Theorem 1, we let $U_1^n(W_{1a},W_{1b},W_{1k},W')$ and $U_2^n(W_{2a},W_{2b},W_{2k},W'')$ be the codewords selected by Users 1 and 2, where $(W_{1k},W_{2k})$ eventually forms another partial key. Finally, the secret key is chosen to be $(W_{1k},W_{2k},K_z)$. 
				
				Below we show that with the appropriate size of the codebook above, the resulting key leakage rate is negligible. 
				
				\textit{Key leakage analysis}: Let $W_1=(W_{1a},W_{1b})$ and $W_2=(W_{2a},W_{2b})$.  The key leakage averaged over all randomly chosen codebooks can be bounded as follows:
				\begin{align*}
				&I(K_1,K_2;W_1,W_2)
				=I(W_{1k},W_{2k},K_z;W_1,W_2)\\
				&\leq H(W_{1k},W_{2k},K_z) -I(W_{1k},W_{2k},K_z;X^n,Y^n,Z^n|W_1,W_2)\\
				&\leq H(W_{1k},W_{2k},K_z)-H(X^n,Y^n,Z^n)+H(W_1,W_2)
				+H(X^n,Y^n,Z^n|W_1,W_2,W_{1k},W_{2k},K_z)\\
				&\overset{(a)}{\leq} n(I(Y,Z;U_1)+I(X,Z;U_2)-I(U_1;U_2)-\delta_{\epsilon})+nR_{k,z} - nH(X,Y,Z) 
				+ n(I(X;U_1|Y,Z)+2\delta_{\epsilon})\\&\qquad  + n(I(Y;U_2|X,Z)+2\delta_{\epsilon}) +n(H(X,Y,Z|U_1,U_2)-R_{k,z}+\delta_{\epsilon}') \\
				&\overset{(b)}{\leq} n\delta_{\epsilon}'',
				\end{align*}
				where in $(a)$, we use the property of the codebook, and Lemma~\ref{lemma:entropy_bound_converse} below, and $(b)$ follows from the Markov chain $U_2-(Y,Z)-(X,U_1)$.
				
				\begin{lemma}\label{lemma:entropy_bound_converse}
					From the codebook generation given above, if $\mathrm{Pr}((X^n,Y^n,Z^n,U_1^n,U_2^n) \in \mathcal{T}_{\epsilon}^{(n)}(X,Y,Z,U_1,U_2)) \rightarrow 1$ as $n \rightarrow \infty$, we have that $H(X^n,Y^n,Z^n|W_1,W_2,W_{1k},W_{2k},K_z) \leq n(H(X,Y,Z|U_1,U_2)-R_{k,z}+\delta_{\epsilon}')$.
				\end{lemma}
				\begin{IEEEproof}
					We consider the following bound:
					\begin{align*}
					&H(X^n,Y^n,Z^n|W_1,W_2,W_{1k},W_{2k},K_z)\\
					&= H(X^n,Y^n,Z^n|W_1,W_2,W_{1k},W_{2k},K_z,W',W'')
					+I(X^n,Y^n,Z^n;W',W''|W_1,W_2,W_{1k},W_{2k},K_z)\\
					&\leq H(X^n,Y^n,Z^n|W_1,W_2,W_{1k},W_{2k},W',W'',K_z)+H(W',W''|W_1,W_2,W_{1k},W_{2k})\\
					&\leq H(Z^n|W_1,W_2,W_{1k},W_{2k},W',W'',K_z)+H(X^n,Y^n|W_1,W_2,W_{1k},W_{2k},W',W'',Z^n)\\
					&\qquad+H(W',W''|W_1,W_2,W_{1k},W_{2k})\\
					&\leq n(H(Z|U_1,U_2)-R_{k,z}+\delta_{\epsilon})  + n(H(X,Y|U_1,U_2,Z)+\delta_{\epsilon}) + n\epsilon_n\\
					&= n(H(X,Y,Z|U_1,U_2)-R_{k,z}+\delta_{\epsilon}'),
					\end{align*}
					where the last inequality follows from the bound $H(Z^n|W_1,W_2,W_{1k},W_{2k},W',W'',K_z) \leq n(H(Z|U_1,U_2)-R_{k,z}+\delta_{\epsilon})$ (see e.g., \cite[Lemma 3]{ckOSSC13}) which holds for $R_{k,z} \leq H(Z|U_1,U_2)-\delta_{\epsilon}$,  from the bound  $H(X^n,Y^n|W_1,W_2,W_{1k},W_{2k},W',W'',Z^n)$ $\leq n(H(X,Y|U_1,U_2,Z)+\delta_{\epsilon})$ which follows from properties of jointly typical sequences (a similar proof can be found in Appendix \ref{app:proof_bound_conditional_entropy} below), and from Fano's inequality $H(W',W''|W_1,W_2,W_{1k},W_{2k}) \leq n\epsilon_{n}$ which holds since given the codebook and $(W_1,W_2,W_{1k},W_{2k})$, the codewords $(U_1^n,U_2^n)$ and thus $(W',W'')$ can be decoded with high probability (the mutual packing lemma \cite{ekNIT11}).
				\end{IEEEproof}
				
				Next, we consider the key rate analysis.
				\begin{align*}
				&H(K_1) = H(W_{1k},W_{2k},K_z)\\
				&=H(W_{1k},W_{2k}) + H(K_z|W_{1k},W_{2k})\\
				&\geq H(W_{1k},W_{2k}) + I(K_z;X^n,Y^n,Z^n|W_{1k},W_{2k})\\
				&\geq H(X^n,Y^n,Z^n) - H(X^n,Y^n,Z^n|W_{1k},W_{2k},K_z)\\
				&= H(X^n,Y^n,Z^n) - H(X^n,Y^n,Z^n|W_1,W_2,W_{1k},W_{2k},K_z) -I(X^n,Y^n,Z^n;W_1,W_2|W_{1k},W_{2k},K_z)\\
				&\overset{(a)}{\geq} nH(X,Y,Z) -n(H(X,Y,Z|U_1,U_2)-R_{k,z}+\delta_{\epsilon}')
				-H(W_1)-H(W_2)\\
				&\overset{(b)}{\geq} n(I(X,Y,Z;U_1,U_2) - I(X;U_1|Y,Z) -I(Y;U_2|X,Z)  + H(Z|U_1,U_2)-\delta_{\epsilon}'')\\
				&\overset{(c)}{=} n(I(Y,Z;U_1)+I(X,Z;U_2)-I(U_1;U_2)+H(Z|U_1,U_2)-\delta_{\epsilon}''),
				\end{align*}
				where $(a)$ follows from Lemma \ref{lemma:entropy_bound_converse},  $(b)$ follows from the codebook generation, and $(c)$ follows from the Markov chain $U_2-(Y,Z)-(X,U_1)$.
				
				\section{Proof of Theorem \ref{theorem:region_special_case}}\label{app:proof_trusted_relay}
				\emph{Sketch of achievability}: For codebook generation, randomly and independently generate codewords $v^n(w_c,w')$ each $\sim P_{V}$, where $w_c \in [1:2^{n(\max\{I(X;V|Y),I(Y;V|X)\}+2\delta_{\epsilon})}]$ and $w' \in [1:2^{n(\min\{I(X;V),I(Y;V)\}-\delta_{\epsilon})}]$. For encoding, given $(x^n,y^n)$, the relay finds a jointly typical codeword $v^n$ and sends the corresponding bin index  $w_c$ to the users. The secret key is chosen as $w'$. With high probability, both users, given $w_c$ and its source, can decode $v^n$, and thus $w'$ correctly. The key leakage averaged over all randomly chosen codebooks can be bounded as follows:
				\begin{align*}
				I(K_1,K_2;W_c)  &= I(W';W_c)\\
				&\leq H(W_c)-I(W_c;X^n,Y^n|W')\\
				&\leq H(W_c)-H(X^n,Y^n) + H(W') + H(X^n,Y^n|W_c,W')\\
				&\overset{(a)}{\leq} n(I(X,Y;V)+\delta_{\epsilon}) -nH(X,Y) + n(H(X,Y|V)+\delta_{\epsilon})\\
				&= n\delta_{\epsilon}',
				\end{align*} 
				where $(a)$ follows from the codebook generation and the bound $H(X^n,Y^n|W_c,W') \leq n(H(X,Y|V)+\delta_{\epsilon})$ (which can be shown similarly as in the proof in Appendix \ref{app:proof_bound_conditional_entropy} below).
				
				Then we have that 
				\begin{align*}
				H(K_1)&= H(W') 
				\geq H(W'|W_c)\\
				&= H(W_c,W')-H(W_c)\\
				&\overset{(a)}{\geq}  H(V^n)-H(W_c)\\
				&\overset{(b)}{\geq} n(I(X,Y;V)-\delta_{\epsilon}) -  n(\max\{I(X;V|Y),I(Y;V|X)\}+2\delta_{\epsilon})\\
				&= n(\min\{I(X;V),I(Y;V)\}-\delta_{\epsilon}')\\
				&\geq n(R_k -\delta_{\epsilon}')
				\end{align*}
				if $R_k \leq \min\{I(X;V),I(Y;V)\}$, where $(a)$ follows from the fact that given the codebook  $V^n$ is a function of $(W_c,W')$, and $(b)$ follows from the encoding processes where $\mathrm{Pr}(V^n=v^n) \leq 2^{-n(I(X,Y;V)-\delta_{\epsilon})}$ and  from the codebook generation.
				\vspace{0.3cm}
				
				\emph{Converse}: Let $V_i =(W_c,K_1,K_2,X^{i-1},Y^{i-1})$. For any achievable $(R_c,R_k)$, it follows that
				\begin{align*}
				nR_c &\geq H(W_c)\\& \geq H(W_c|Y^n)-H(W_c,K_1,K_2|X^n,Y^n)\\
				&\overset{(a)}{\geq} I(W_c,K_1,K_2;X^n|Y^n)-n\epsilon_{n}\\
				&\overset{(b)}{\geq} \sum_{i=1}^n I(X_i;V_i|Y_i)-n\epsilon_{n},
				\end{align*}
				where $(a)$ follows from the fact that $K_2$ is a function of $(W_c,Y^n)$ and Fano's inequality $H(K_1|W_c,Y^n)\leq n\epsilon_{n}$, and $(b)$ follows from the definition of $V_i$.
				
				By symmetry, it also follows that $nR_c \geq \sum_{i=1}^n I(Y_i;V_i|X_i)-n\epsilon_{n}$.
				
				For the key rate, it follows that
				\begin{align*}
				n R_k &\leq H(K_1) 
				\leq H(K_1,K_2)\\
				&\overset{(a)}{\leq} H(K_1,K_2|W_c) +n\delta_{n}\\
				&\overset{(b)}{\leq} I(K_1,K_2;Y^n|W_c) +n\delta_{n} +n\epsilon_{n}\\
				&\overset{(c)}{\leq} \sum_{i=1}^n I(Y_i;V_i)+n\delta_{n} +n\epsilon_{n},
				\end{align*}
				where $(a)$ follows from the key leakage constraint, $(b)$ follows from the fact that $K_2$ is a function of $(W_c,Y^n)$ and Fano's inequality $H(K_1|W_c,Y^n)\leq n\epsilon_{n}$, and $(c)$ follows from the definition of $V_i$.
				
				Similarly, by symmetry, it also follows that $nR_k \leq \sum_{i=1}^n I(X_i;V_i)+n\delta_{n} +n\epsilon_{n}$. 	The proof ends by standard steps of introducing a time-sharing random variable and letting $n \rightarrow \infty$.

				\section{Proof of Bound on Conditional Entropy $H(X^n,Y^n|W_1,W_2,W_{1k},W_{2k},W',W'') \leq n(H(X,Y|U_1,U_2)+\delta_{\epsilon})$}\label{app:proof_bound_conditional_entropy}
				Let $E$ be a binary random variable taking value $0$ if $(X^n,Y^n,U_1^n(W_1,W_{1k},W'),U_2^n(W_2,W_{2k},W'')) \in \mathcal{T}_{\epsilon}^{(n)}$, and $1$ otherwise. Since $(X^n,Y^n,U_1^n,U_2^n) \in \mathcal{T}_{\epsilon}^{(n)}$ with high probability from the encoding process, then $\mathrm{Pr}(E=1) \leq \delta_{\epsilon}$. We have that
				\begin{align*}
				&H(X^n,Y^n|W_1,W_2,W_{1k},W_{2k},W',W'')\\
				&\overset{(a)}{\leq} H(X^n,Y^n,E|U_1^n(W_1,W_{1k},W'),U_2^n(W_2,W_{2k},W''))\\
				&\leq H(X^n,Y^n|U_1^n,U_2^n,E) + H(E)\\
				&= \mathrm{Pr}(E=0) H(X^n,Y^n|U_1^n,U_2^n,E=0)  + \mathrm{Pr}(E=1) H(X^n,Y^n|U_1^n,U_2^n,E=1) + H(E)\\
				&\overset{(b)}{\leq}H(X^n,Y^n|U_1^n,U_2^n,E=0) +\delta_{\epsilon} H(X^n,Y^n)+ h(\delta_{\epsilon}) \\
				&\leq H(X^n,Y^n|U_1^n,U_2^n,E=0) + n\delta_{\epsilon} \log|\mathcal{X}||\mathcal{Y}|  + h(\delta_{\epsilon})\\
				&\leq \sum_{(u_1^n,u_2^n) \in \mathcal{T}_{\epsilon}^{(n)}} p(u_1^n,u_2^n|E=0) \cdot  H(X^n,Y^n|U_1^n=u_1^n,U_2^n=u_2^n,E=0)  + n\delta_{\epsilon}'\\
				&\overset{(c)}{\leq} \sum_{(u_1^n,u_2^n) \in \mathcal{T}_{\epsilon}^{(n)}} p(u_1^n,u_2^n|E=0)  \log|\mathcal{T}_{\epsilon}^{(n)}(X,Y|u_1^n,u_2^n)|  + n\delta_{\epsilon}'\\
				&\overset{(d)}{\leq} n(H(X,Y|U_1,U_2)+\delta_{\epsilon}'),
				\end{align*}
				where step $(a)$ follows from the fact that given the codebook, $U_1^n$ and $U_2^n$ are functions of $(W_1,W_{1k},W')$ and $(W_2,W_{2k},W'')$, $(b)$ follows from $\mathrm{Pr}(E=1) \leq \delta_{\epsilon}$ where $h(\cdot)$ is the binary entropy function, and $(c)$ and $(d)$ follow from the properties of jointly typical set \cite{ekNIT11} with $\delta_{\epsilon}, \delta_{\epsilon}' \rightarrow 0$ as $\epsilon \rightarrow 0$, and $\epsilon \rightarrow 0$ as $n \rightarrow \infty$. 
				
				We note that similar proofs can be obtained for other conditional entropy bounds of this form, e.g., \[H(X^n|W_{1k},W_{2k},W_1,W_2,W',W'') \leq n(H(X|U_1,U_2)+\delta_{\epsilon})\] and \[H(Y^n|W_2,W_{2k},W'',X^n) \leq n(H(Y|U_2,X)+\delta_{\epsilon})\] which appear in achievability proof in the paper.

				\bibliographystyle{IEEEtran}
				\bibliography{IEEEabrv,ITW2016_arxiv}
\end{document}